# 3D shaping of electron beams using amplitude masks

Roy Shiloh[1]* and Ady Arie[1]

[1]School of Electrical Engineering, Fleischman Faculty of Engineering, Tel Aviv University, Tel Aviv, Israel.

*royshilo@post.tau.ac.il

ABSTRACT: Shaping the electron wavefunction in three dimensions may prove to be an indispensable tool for research involving atomic-sized particle trapping, manipulation, and synthesis. We utilize computer-generated holograms to sculpt electron wavefunctions in a standard transmission electron microscope in 3D, and demonstrate the formation of electron beams exhibiting high intensity along specific trajectories as well as shaping the beam into a 3D lattice of hot-spots. The concepts presented here are similar to those used in light optics for trapping and tweezing of particles, but at atomic scale resolutions.

## 1. INTRODUCTION

The spatial manipulation of the electron wave function has attracted much attention in recent years. Essential techniques such as computer generated holography and special beams carrying orbital angular momentum (vortex beams [1–5]), beams exhibiting self-acceleration [6] as well as non-diffracting beams [7] have been adapted from light optics and applied to electron beams, thereby enabling to shape electron beams with dimensions that are comparable with the size of atoms, i.e. at the sub-nanometer scale. However, up until now, all these demonstrations were spatially limited to two-dimensional shaping of the wavefunction in a particular plane, usually the focal plane of the lens that forms the desired shape. In this study we show, for the first time, how the electron beam in a standard transmission electron microscope (TEM) is spatially shaped nearly arbitrarily in three-dimensions (3D). We give examples of an intensity distribution shaped as an Archimedean spiral, and a face-centered cubic-like lattice of high intensity, focused spots. The significance of this novel tool is in its applications to particle trapping and manipulation in a volume, and the possibility of controlled, assisted-synthesis of single particles and aggregates.

The spatial manipulation of the electron wave function has taken a leap forward recently, with nano machinery such as the focused ion beam playing a pivotal role in fabrication. Thin silicon nitride membranes are widely used as substrates for electron phase manipulation, mainly due to their mechanical robustness, low scattering, and commercial availability. The authors have previously shown [8], that such phase masks may be engineered as computer generated holograms to create a desired, nearly-arbitrary intensity pattern in the diffraction plane, using an inverse Fourier transform algorithm such as Gerchberg-Saxton's [9,10].

In this work, we present for the first time, the nearly-arbitrary manipulation of the electron wave function in 3D, with the purpose of introducing a catalyst for research in particle trapping, manipulation, and even material synthesis, using the TEM's extraordinary resolution. Some measure of control over the rotation of particles in the TEM has already been shown [11,12], while particle synthesis using electron beam irradiation (not necessarily in the TEM) has been explored [13–15]. In his revolutionary 1970's paper, Ashkin demonstrated [16] trapping and manipulation of particles by the radiation pressure of laser light. Since then, this field has evolved enormously and includes applications in biology, chemistry, and physics [17,18]. Shaped electron tweezers and manipulators do not only have the potential advantage of atomic resolution, but also the added value of being formed of charged particles, which may provide interesting interaction dynamics with different neutral, charged, and magnetic particles. For example, particles sensitive to electric or magnetic fields may be sorted using correctly shaped electron beams. As such, rather than shaping the beam to reform in the diffraction plane only, we design our holograms to yield an intensity pattern that follows a predetermined curve while propagating along the TEM column, or singular "hotspots", or traps, in a lattice configuration. We've chosen two examples, based on methods developed in light-optics, to demonstrate this capability.

## 2. BINARY AMPLITUDE MASKS

Different authors have previously [5,6,19,20] used binary amplitude masks to generate off-axis vortex beams, and blazed gratings as phase masks to optimize off-axis diffraction efficiency [21–23]. In light optics, binary amplitude masks, or binary computer generated holograms, were first introduced by Brown & Lohmann [24]. These act as opaque slides with carefully designed holes, completely blocking or transmitting light, in such a way as to form a desired intensity pattern in the far-field. Similarly, in electron optics, amplitude masks have been fabricated by milling through thin platinum foils, or coating a silicon nitride membrane with high-Z metals and milling through them. It is important to remember, however, that in contrast to light optics, where photons may indeed be completely blocked by such a mask, it would take several tens or hundreds of microns of a high-Z metal to actually stop an electron with a typical energy of 200keV in a standard transmission electron microscope. Moreover, such hard-stopped electrons may generate extensive damage and heating to the film, depending on its make. Instead, thin metal films or coatings yield high-angle scattering related to lattice constants, which is the basic mechanism for all amplitude masks considered above. In electron beam shaping, we usually only investigate much smaller angles, as we are limited by our fabrication capabilities, and in that sense electron amplitude masks operate in accordance with their light-optics counterparts. In our fabrication process, we employ, for the first time to the best of our knowledge, a fully standard electron beam lithography scheme, complete with resist, metal coating, and liftoff, on thin silicon-nitride membranes, as elaborated on in Appendix A. Our experimental setup consists of such masks placed in a standard TEM sample holder, where we observe the resulting diffraction pattern and the planes in its vicinity, as is illustrated in Fig. 1. The figure shows the optical setup depicting the zero and first diffraction orders, where the latter are shaped as Archimedean spirals.

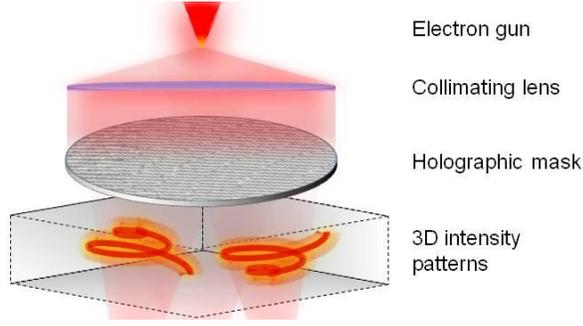

Fig. 1: Illustration of the experimental setup in the TEM for shaping electron beams in 3D. (a) The illumination system creates a collimated beam which passes through a holographic amplitude mask. (b) As the electron beam propagates away from the mask, high-intensity 3D curves and patterns are formed according to the design encoded into the mask and imprinted onto the electron wavefunction.

We utilize Lee's encoding method to design computer generated binary off-axis holograms which reconstruct the desired wavefront in the first diffraction order [25]. This simple analytic method encodes both amplitude and phase of a desired complex wave $U$ into a binary amplitude mask using the simple formula:

$$MASK(x,y) = \frac{1}{2}[1 + sign\{\cos[2\pi f_c x + \phi(x,y)] - \cos[\pi a(x,y)]\}], \quad (1)$$

where $f_c = 1/\Lambda$ is a carrier frequency designed to reconstruct the hologram off-axis, and determined from a grating of period $\Lambda$, and $\phi(x,y)$ and $\sin \pi a(x,y) = A(x,y)$ are related to the Fourier transform of the object wave to be recorded. In the above case, this amounts to $FT\{H\} = h(x,y) = A(x,y)\exp[i\phi(x,y)]$. The wavefront $H(\vec{r}',t)$ is reconstructed in the first diffraction order. For the purpose of lowering the potential overlap of the extended information in the diffraction orders, we modulate along a diagonal grating, i.e. replace $2\pi f_c x$ by $2\pi f_c (x+y)$ above.

## 3. HIGH-INTENSITY CURVES: THE ARCHIMEDEAN SPIRAL

Let us now show how an electron beam can be formed to follow a specific trajectory in three-dimensional space. As an example, we consider the Archimedean spiral. The spiral is parametrically defined as,

$$\vec{c}_3(t) = (x_0(t), y_0(t), z_0(t)) = \left(-tR\cos(\omega t), -tR\sin(\omega t), sR\left[\frac{1}{2} - \sqrt{1-t^2}\right]\right), \quad (2)$$

where $t$ is the normalized parameter of evolution of the spiral in the propagation direction ($z$), $R$ is the maximal radius in the transverse direction ($x,y$), and $\omega$ is the angular frequency of rotation of the spiral. The relation between these values and the actual measured values in the experiment depends on the focal length and numerical aperture of the optical system. This beam could be used to channel particles along a specified curve. A similar work on solenoid beams in light optics [26] has shown the ability to pull, hold, or push particles.

For designing the computer-generated hologram, we follow the methods introduced in light optics by Abramochkin and Volostnikov [27,28] and Rodrigo et. al. [29]. In this non-iterative technique, the beam is designed to have a high intensity distribution which follows a designed path through continuous planes along the propagation direction. The beam, parameterized by $t$, is given by

$$H(\vec{r}',t) = \frac{1}{L}\int_0^T \varphi(\vec{r}',t)\Phi(\vec{r}',t)\left|\frac{dc_3}{dt}\right|dt, \quad (3)$$

where $L$, $\varphi(\vec{r}',t)$, and $\Phi(\vec{r}',t)$ are as defined in [29], and $|\vec{c}_3(t)|$ is the $L^2$ norm of the vector, $|\vec{c}_3(t)| = \sqrt{x_0^2(t) + y_0^2(t) + z_0^2(t)}$. To avoid confusion with notation, here $\vec{r}' = (x',y',z')$ are the spatial coordinates of the Archimedean spiral near the diffraction plane, while in the following $(x,y,z)$ are the spatial coordinates in the (image) plane of the amplitude mask. For this demonstration, we choose $t \in [0, 0.75]$, and normalized values for $R = 40$ and $\omega = 20$, the final scaling of which depends on the optical setup of the microscope (see Appendix B). In Fig. 2a, a depiction of this curve is demonstrated in 3D, and in Fig. 2b-h we present measured images of the first diffraction order taken at different planes along the propagation axis, where it is seen that the highest intensity (focused) segment of the spiral continuously changes in accordance with the designed curve. A real-time recording of the positive and negative first diffraction orders at different distances from the focal plane is also available online.

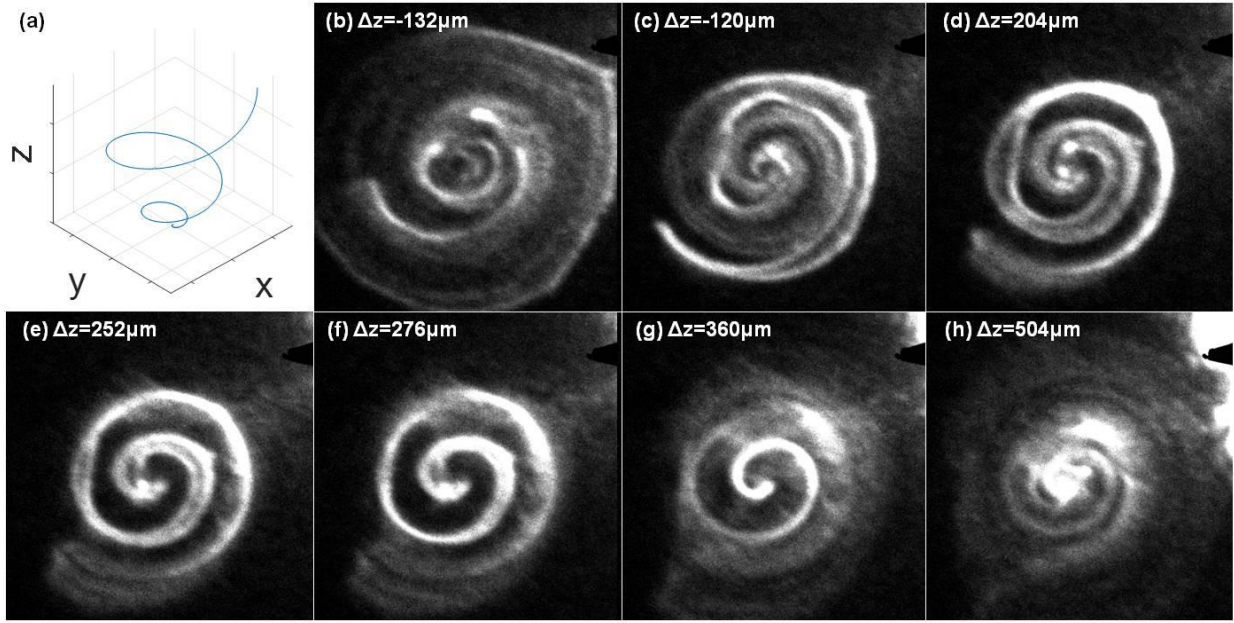

Fig. 2: The Archimedean spiral. (a) 3D plot of the spiral. (b) Measurements taken at different planes along the propagation axis. $\Delta z$ is measured with respect to the (focused) diffraction plane. The diameter of the spiral is roughly $455 \mu m$. Contrast and brightness have been altered for visibility.

## 4. VOLUME TRAPS: FACE-CENTERED CUBIC-LIKE LATTICE

Our second demonstration, also motivated by particle manipulation and synthesis applications, explores the concept of generating particle traps arranged as a face-centered cubic Bravais-like lattice. Such traps could be used to pin down nano particles for observation [30], or be used to generate high-intensity patterns to spatially control particle synthesis [13–15]. With a slight change of the optical setup, such traps could be fabricated and measured even in the atomic scale. Here we employ a formalism based on inverted Gabor holographic technique, proposed by Latychevskaya et. al [31]. In this elegant formulation, a point scatterer (black pixel) is simulated using a standard beam propagation method. The scatterer would ideally emit a spherical wavefront, which is apparent in the simulation. Additional scatterers are placed in other parallel planes where desired, and eventually the simulated wavefront is inverted such that all scatterers act now as focal points of spherical waves in superposition. A similar approach was studied by Lohmann in 1969 [32]. The wavefront is then inverse-Fourier transformed so as to find the required wavefront in the hologram plane. In our case, the longitudinal, optical depth of the hologram $\Delta z$ is much smaller than the focal length of the diffraction lens, hence this lens provides most of the focusing power and the mask only adds small perturbations. Once the electron beam passes through the mask, its propagation dynamics in the transverse and longitudinal directions are determined. In Fig. 3a, we present the reconstructed 3D model [33] from a focal series measurement. Fig. 3b shows an overview of the diffraction plane with the zero order blocked and first diffraction orders (bright blobs) showing at the edges. Intensity-wise, the holographic reconstruction and its conjugate appear and behave identically in this case, since the lattice is symmetrically positioned on both sides (along z) of the diffraction plane. Fig. 3c-d shows two different planes of the lattice at focus; there, the electron count in the traps is about 700 times larger than the background (in an exposure time of 250ms). Further optimization of the diffracted intensity can be done by using different fabrication methods, e.g. using phase hologram [8] instead of the amplitude hologram and more sophisticated holographic encoding techniques. Additional discussion on the scales of our measurements is provided in Appendix B.

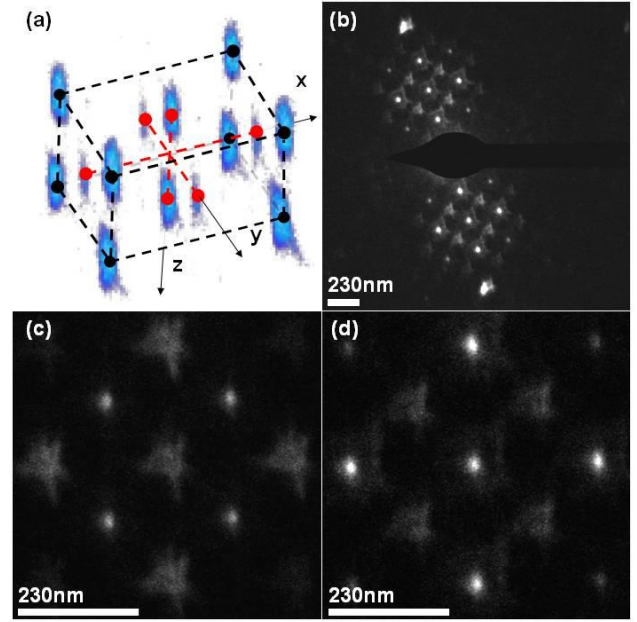

Fig. 3: Generation of a face-centered cubic intensity pattern. (a) Reconstructed 3D model through a focal series. (b) Diffraction pattern showing the first diffraction orders (zero order is blocked). (c) Slice from the middle of the lattice. (d) Slice from the bottom of the lattice. Contrast and brightness have been altered for visibility.

## 5. CONCLUSION

In conclusion, we have demonstrated, for the first time in electron optics, the nearly-arbitrary manipulation of the electron wave function in 3D, using computer generated holograms. Our holograms are based on amplitude masks, the fabrication of which fully follows a standard electron beam lithography process. The mask consists of a well-ordered array of gold islands, the advantage of which is conductivity and high-Z scattering. The ability to manipulate electron beams in 3D heralds important possibilities for research in electron microscopy: using a liquid cell holder, one could envision spatially and temporally trapping particles for careful scrutiny, inducing and controlling the rotation of a single particle, and the interesting concept of material synthesis by trapping single atoms in a shaped, 3D electron matrix, the structure and scale of which could be dynamically controlled. The same technique we show here can be used to form electron beams with more complex shapes, enabling for example the generation of knotted electron wave-function [34] and thus enabling to study its interaction with different materials.

## APPENDIX A    FABRICATION AND MEASUREMENT PROCESSES

In our fabrication process, we employ a fully standard electron beam lithography scheme, complete with resist, metal coating, and liftoff, on thin silicon-nitride membranes. First, the membrane is spin-coated with PMMA 950K electron-beam resist, to reach a layer of about 200nm. The membrane is baked at 130 degrees Celsius for 5 minutes on a hot plate, and inserted into a Raith 150 e-beam lithography system. Then, the encoded hologram is transferred into the resist as a matrix of dots with a period of $\delta = 120nm$, using a dose of 0.01pC. The sample is developed in a 1:3 MIBK:IPA solution for 60 seconds, and later coated with 5nm Cr and 25nm Au in an e-beam evaporator. Then, lift-off is promptly achieved in an NMP solution, leaving tiny gold islands which collectively diffract the electron wave into many diffraction orders, along with the desired first order. It may be argued that this gold layer is insufficient to act as a dominant amplitude-only scatterer, and that it also behaves as a strong phase object. However, while this may be true, our measurements presented in this article deem the argument irrelevant for the fulfillment of the purpose of our masks. Figure A.1 shows an example of a mask (a) and its diffraction pattern (b) at focus; the diffraction orders are recognizable from the dot matrix basis, where the first diffraction orders are in angle $\theta = 20.8 \mu rad$, related to the period $\delta = 120nm$ in the mask. Around them, the 3D spiral modulation is visible in diffraction orders with an angle related to the $\Lambda = 4\delta$ carrier period in the mask.

Our measurements throughout this work are performed as follows. We place the amplitude mask on a standard sample holder, then use low magnification mode to locate the 90um mask. Additionally, we insert a 100um objective aperture, which in this operating mode is optically close enough to the image plane to act as a selected-area aperture. We then go to low-angle diffraction mode and focus the unscattered, zero order beam, using Condenser 2. Now, the objective lens may be used to observe the evolution of the diffraction pattern in different planes along the propagation ($z$) axis, on the camera.

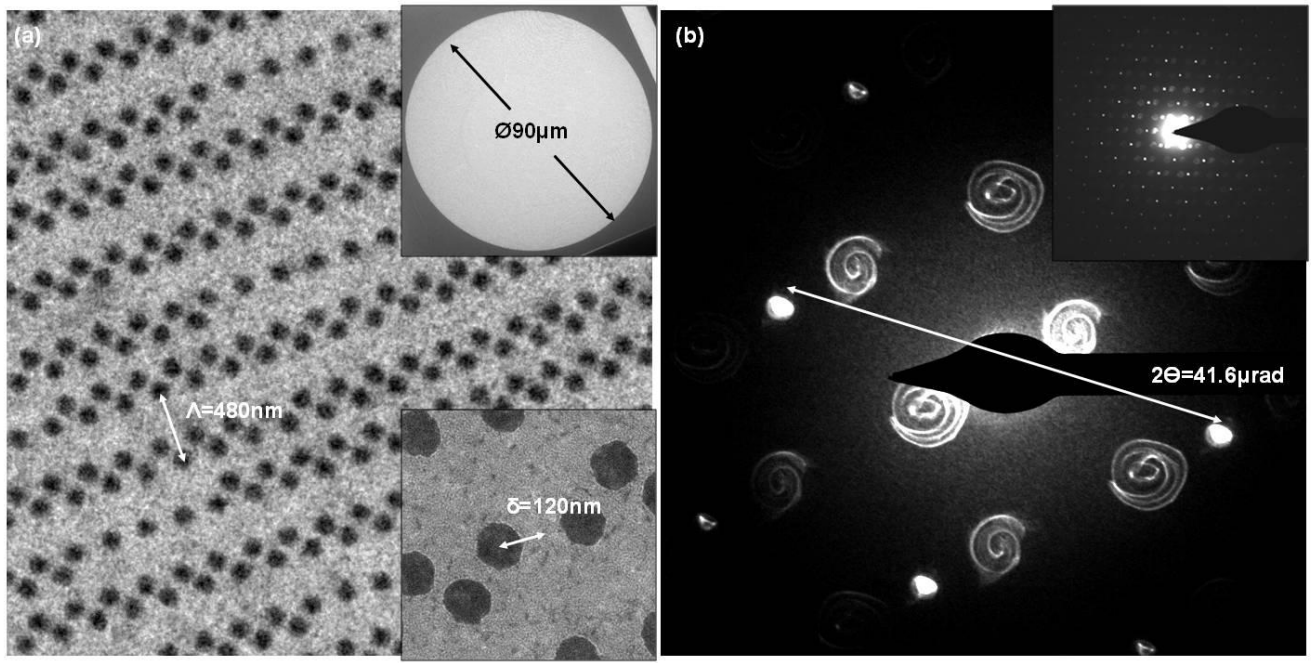

Fig. A.1: Example mask and diffraction pattern. (a) Gold dot matrix on SiN membrane, with a period of $\delta = 120nm$. Insets show the full $90 \mu m$ mask and a high-magnification image of the gold nano islands. (b) Diffraction pattern showing the Archimedean spiral, relating the diffraction orders to the mask's periodicity. Inset shows multiple diffraction orders in low angle diffraction. Contrast and brightness have been altered for visibility.

In general, additional orders may be filtered, if so desired, by placing a selecting-area aperture in an image of the diffraction plane. These could be used for an application requiring spatial multiplications, though in such a case a better option would be to resort to Dammann gratings [35,36]. It is worth noting that when dealing with 3D shaping and diffractive elements in general, such multiplications may also manifest in the longitudinal direction (see e.g. [37]). For the goal of this paper, we are only interested in the volume containing our 3D-shaped beam, which we will exclusively investigate next.

## APPENDIX B  AXES CALIBRATION

Our electron microscope, the Tecnai F20 operating at 200kV ($\lambda \approx 2.5 pm$), is equipped with a Gatan Oneview camera having a 4K sensor. The mask is illuminated with a near-parallel beam, its diameter nearing $D = 100 \mu m$, and we assume a focal length of $f = 10 cm$ for the diffraction lens (which acts as the imaging lens in this mode). Thus, the numerical aperture is calculated to be $\beta = D/2f = 500 \mu rad$. The first diffraction order is separated by an angle $\theta = \lambda/\delta \approx 21 \mu rad$, where $1/f_c = \delta$ is the carrier modulation period. To directly measure the magnification to the CCD, we first calculate the theoretical, unmagnified distance between the zero and first diffraction order: $x = \theta f \approx 2.1 \mu m$. Given the physical pixel size on the CCD, $px = 15 \mu m$, and the number of pixels actually measured, we deduce the magnification to be $M \approx 21500$.

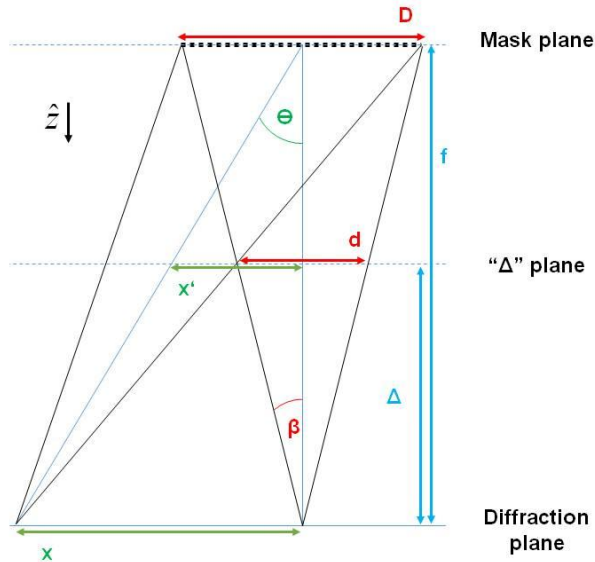

Figure B.1: Geometrical representation of the optical system and associated quantities used for axes calibration. "$\Delta$-plane" refers to the plane at a distance $\Delta$ from the focused diffraction pattern where the zero and first order begin to overlap. Note: figure is not to scale.

By simple geometrical calculation (see Figure B.1), the distance from the focal plane in which the zero and first order begin to overlap can be expressed in several equivalent ways:

$$\Delta = \frac{f}{1+2\beta/\theta} = \frac{f}{1+D/\theta f} = \frac{D}{2\beta(1+2\beta/\theta)}. \quad (4)$$

In our case, we find that $\Delta = 2.06 mm$. Since during measurement we only slightly change the objective lens (less than 0.1% in the $\Delta$-plane), the magnification, which strongly depends only on the lenses below the diffraction lens, is constant. We have corroborated this assumption by calculating the zero order beam's diameter: $d = D\Delta/f = 2.06 \mu m$, and measuring it directly on the CCD using its pixel size. Considering the magnification above, this results in $2.15 \mu m$, where at least some of the deviation may be attributed to the our basic assumption on the diffraction lens's focal length.

Using the Oneview camera, we were able to record real-time movies of the focal series (available online). This was achieved by changing the objective lens at constant speed (controlled by a Free Lens software), and taking images at 4 frames per second, with integration time of 0.25sec. This was done from the $\Delta$-plane to the inverse $\Delta$-plane, i.e. below the diffraction plane. For the chosen speed and integration time, we recorded 172 frames between the $\Delta$-plane and the diffraction plane, yielding a longitudinal resolution of $\delta z = \Delta/172 \approx 12 \mu m$. Finally, the transverse resolution in our experimental system is simply deduced by $\delta x = px/M \approx 700 pm$. It is important to note that, while the maximal recordable size on the CCD is thus $4096 \delta x \approx 2.87 \mu m$, the maximal spatial bandwidth in the diffraction plane is restricted by fabrication through $\delta$ and our encoding method's carrier frequency through $\Lambda$. An estimate may be given by demanding that diffraction orders do not overlap, leading to the relaxed limitation $spBW < 1.05 \mu m$, which amounts to less than 37% of the CCD. From these results, it is obvious that by using low magnification mode it would be difficult to fabricate and measure 3D structures of equal scales in all axes. A possible solution to this problem is to transfer the mask to the C2 aperture, where much higher magnifications of the diffraction pattern and larger numerical apertures may be achieved.

## APPENDIX C  REAL-TIME VIDEO RECORDINGS

Movie C.1 (online): a real-time recording (continuous focal series) of the three-dimensional Archimedean spiral, as measured in our TEM. The movie shows both image and its conjugate spiral, swirling in opposite (z) directions. Besides being asymmetric in z, The spirals are slightly shifted from the position of the diffraction plane (in z), and thus we see them in focus at different planes relative to one another. Contrast and brightness have been altered for visibility.

Movie C.2 (online): a real-time recording (continuous focal series) of one diffraction order of the FCC-like lattice pattern. The non-circular behavior of the spots when out of focus is a result of charged contamination on the edges of our objective aperture. Contrast and brightness have been altered to better visualize the converging and diverging beams; the electron counts in focused spots is about 700

times that of the background (in an exposure time of 250ms).

**FUNDING**

The work was supported by the Israel Science Foundation, grant no. 1310/13, and by the German-Israeli Project cooperation (DIP).